\documentclass{svjour3}                    
\smartqed 
 \usepackage{graphicx}
\usepackage{fix-cm}
\usepackage{amsmath}
\def\n{\nonumber}
\def\p{\partial}
\def\be{\begin{equation}}
\def\ee{\end{equation}}
\def\d{{\rm d}}
\def\ba{\begin{eqnarray}}
\def\ea{\end{eqnarray}}
\def\ve{\varepsilon}

\journalname{International Journal of Theoretical Physics}

\begin{document}

\title{Applications of Lie Symmetries to Higher Dimensional Gravitating Fluids}

\author{A. M. Msomi      \and K. S Govinder \and
        S. D. Maharaj}

\institute{
A. M. Msomi \and K. S. Govinder        \and S. D. Maharaj \at
Astrophysics and Cosmology Research Unit,
 School of Mathematical Sciences, Private Bag X54001,
 University of KwaZulu-Natal, Durban 4000, South Africa
\\
\email{maharaj@ukzn.ac.za}  \\ \\
A. M. Msomi\\
Department of Mathematical Sciences,
 Mangosuthu University
of Technology,  P. O. Box 12363, Jacobs  4026, South Africa}

\date{Received: date / Accepted: date}

\maketitle

\begin{abstract}
We consider a radiating shear-free spherically symmetric metric in higher dimensions.
 Several new solutions to the Einstein's equations are found systematically
 using the method of Lie analysis of differential equations. 
 Using the five Lie point symmetries of the fundamental field equation, 
 we obtain either an implicit solution or we can reduce 
 the governing equations to a Riccati equation. We show that
 known solutions of the Einstein equations can produce 
 infinite families of new solutions. Earlier results in four dimensions
 are shown to be special cases of our generalised results.


\end{abstract}

\section{Introduction}
The spherically symmetric radiating spacetimes with vanishing 
shear are important for applications in relativistic astrophysics, radiating 
stars and 
cosmology. In the literature, there exists a large number 
of studies of various models involving gravitational collapse with 
radiative 
processes. Studies modeling relativistic stars show that a 
necessary requirement for these models is that the interior 
radiating spacetime has to be 
matched at the boundary, with the radial pressure being 
nonzero, to the exterior Vaidya radiating spacetime. Krasinski 
\cite{krasinski} pointed out 
the significance of relativistic heat conducting fluids in 
modeling inhomogeneous processes. Some exact solutions in the 
presence of heat 
flow have been developed by Bergmann \cite{berg}, 
Maiti \cite{maiti} and Modak \cite{modak}. 
In considering spherical gravitational collapse, the  appearance of 
singularities and the formation of horizons, Banerjee 
and Chatterjee \cite{banerjee} and Banerjee et al.~\cite{banerjee1} have 
investigated heat 
conducting fluids in higher dimensional cosmological 
models. Davidson and Gurwich \cite{davidson} and Maartens and Koyama 
\cite{maartens} highlighted the 
role of heat flow in gravitational dynamics and perturbations 
in the framework of brane world cosmological models.

 A proper and complete model of a radiating relativistic 
 star requires the presence of heat flux. The result given by 
Santos et al.~\cite{santos}
indicates that the interior spacetime must contain a 
nonzero heat flux so that the matching 
of the interior at the boundary to the exterior 
Vaidya spacetime 
is possible. Models of heat flow in astrophysics have 
been used in gravitational collapse, black hole physics, formation 
of singularities 
and particle production at the stellar surface in four and 
higher dimensions.
Herrera et al.~\cite{herrera1}, Maharaj and Govender 
\cite{maharaj} and Misthry et al.~\cite{misthry}  showed that
heat conducting relativistic radiating 
stars are also useful in the investigation of the cosmic censorship
 hypothesis and in describing collapse with vanishing tidal 
forces. Solutions to the Einstein field equations
 for a shear-free spherically symmetric spacetime with 
a homothetic vector, together with radial heat flux, have been presented 
by Wagh et al.~\cite{wagh}. Analytical solutions 
to the field equations for radiating collapsing spheres in the diffusion 
approximation have
been found by Herrera et al.~\cite{herrera2}. Recent 
examples of radiating stars, with generalised energy momentum tensors,
are given by Herrera et al.~\cite{herrera3} and Pinheiro and Chan \cite{Pinheiro}.

Shear-free fluids, in the presence of heat flux,
 are also important in modeling inhomogeneous 
 cosmological processes. The need for radiating models 
in the formation of 
structure, evolution of voids, the study of singularities, and investigations
of the cosmic censorship hypothesis, has been pointed out by Krasinski \cite{krasinski}. 
Banerjee et al.~\cite{banerjee2} generated a model of a heat conducting
sphere which radiates energy during collapse without the appearance
of a horizon at any stage. This result holds in four dimensions but may be extended
to models in higher dimensions. Banerjee and 
Chatterjee \cite{banerjee} studied  heat conducting fluids  in cosmological models in higher 
dimensions,  and determined that
gravitational collapse is also possible without the appearance of an event horizon. 
The presence of heat flow in brane world models sometimes allows for 
more general 
behaviour than is the case in standard general relativity. 
Govender and Dadhich \cite{mgov} proved that the analogue of the Oppenheimer-Snyder model 
of a collapsing dust permits a radiating brane.

In this paper we analyse the master equation for higher dimensional
radiating fluids studied by Banerjee and Chatterjee \cite{banerjee}  applicable to a $(n
+2)$-dimensional 
spherically symmetric metric. In our analysis, we used the  Lie theory 
of extended groups as a systematic approach to generalise known 
solutions and generate 
new solutions of the same equation. The higher dimensional
radiating model is derived in Sect.~\ref{s2}. In Sect.~\ref{s3}, we give a brief 
outline of the Lie theory. In Sect.~\ref{s4}, we discuss the new 
solutions of the master 
equation that can be found via Lie symmetries
by taking one potential to be a function of the remaining potential. Also 
in this section, we systematically study other group invariant solutions admitted by the 
fundamental 
equation by taking specific ratios
of the potentials. In Sect.~\ref{s5}, we extend known solutions to new solutions 
of the fundamental equation utilising Lie theory. 
Regardless of the 
complexity of the generating function chosen it is possible to find new  
exact solutions; we demonstrate this in two cases. We conclude this 
paper with some 
brief observations about the nature of the new exact solutions in Sect.~\ref{s6}.

\section{Radiating Model \label{s2}}
We consider the shear-free, spherically symmetric line element 
with an exterior $(n+2)$-dimensional manifold  given by
\be
ds^2 = -A^2dt^2+\frac{1}{F^2}\left[dr^2+r^2dX_{n}^2\right]\label{con1}
\ee
where $A=A(t,r)$ and $F=F(t,r)$ and
\be
X_{n}^{2} = d\theta^2_{1}+\sin^2\theta_{1}d\theta^2_{2}+ \dots 
+ \sin^2\theta_{1}\sin^2\theta_{2} \ldots \sin^2\theta_{n-1}d\theta^2_{n}
\ee
The energy momentum tensor for a nonviscous heat conducting fluid is given by
\be
T_{ij} = (\rho+p)v_{i}v_{j}+pg_{ij}+q_{i}v_{j}+q_{j}v_{i}\label{con2}
\ee
where $\rho$ is the energy density of the fluid, $p$ the isotropic 
fluid pressure, $v_{i}$ is the $(n+2)$-velocity and $q_{i}$ 
is the heat flow 
vector. Using equations (\ref{con1}) and (\ref{con2}) we find that 
the nontrivial Einstein field equations in comoving coordinates are
\begin{subequations}
\ba
\rho &=& \frac{n(n+1)F_{t}^2}{2A^2F^2}-\frac{n(n+1)F_{r}^2}{2}+nFF_{rr}+\frac{n^2FF_{r}}{r}\label{3a}\\
p &=& - \frac{nA_{r}FF_{r}}{A}+\frac{nA_{r}F^2}{rA}+\frac{n(n-1)F_{r}^2}{2}-\frac{n(n-1)FF_{r}}{r}\n 
\\&& + \mbox{}\frac{nF_
{tt}}{A^2F}-
\frac{n(n+3)F_{t}^2}{2A^2F^2}-\frac{nA_{t}F_{t}}{A^3F}\label{p1}\\
p &=& \frac{F^2A_{rr}}{A}-(n-1)FF_{rr}+\frac{n(n-1)F_{r}^2}{2}+\frac{(n-1)F^2A_{r}}{rA}\n \\&& - \mbox{}\frac{(n-1)^2FF_{r}}
{r}-\frac{(n-2)
FF_{r}A_{r}}{A}+\frac{nF_{tt}}{A^2F} \n \\
&& -\frac{n(n+3)F_{t}^2}{2A^2F^2}-\frac{nA_{t}F_{t}}{A^3F}\label{p2}\\
q &=& -\frac{nFF_{tr}}{A}+\frac{nF_{t}F_{r}}{A}+\frac{nFF_{t}A_{r}}{A^2}\label{3d}
\ea
\end{subequations}
The isotropy of pressure is given by equations $(\ref{p1})$ and $(\ref{p2})$ together in the form
\be
FA_{xx}+2A_{x}F_{x}-(n-1)AF_{xx} = 0 \label{ben}
\ee
 with $x =r^2$.  Equation (\ref{ben}) is the master equation for the system of higher dimensional
 Einstein field equations with $n \geq 2$. 
In this paper, we 
reduce the order of (\ref{ben}) via Lie analysis in order to find general solutions in higher dimensions.

\section{Lie analysis \label{s3}}
 The symmetry analysis for a system of ordinary differential equations in two dependent variables requires the 
determination of the one--
parameter ($\ve$) Lie group of transformations
\ba
\bar{x} &=& f(x,F,A,\ve) \n \\
\bar{F} &=& g(x,F,A,\ve) \label{2.2a}\\
\bar{A} &=& h(x,F,A,\ve) \n 
\ea
that leaves the solution set of the system invariant. It is difficult to calculate these transformations directly, and as such, we 
must resort to 
approximations via 
\ba
\bar{x} &=& x + \ve \xi(x,F,A) + O(\ve^2) \n \\
\bar{F} &=& F + \ve \eta(x,F,A) + O(\ve^2) \label{2.2} \\
\bar{A} &=& A +\ve \zeta(x,F,A) + O(\ve^2)  \n
\ea
The transformations (\ref{2.2}) can be obtained once we find the  (symmetry) operator  \be  Z = \xi \frac{\p\ }{\p x} + \eta 
\frac{\p\ }{\p F}+ 
\zeta \frac{\p }{\p A} \label{2.4} \ee
which is a set of vector fields. Once these symmetries are determined,  it is possible to regain the finite (global) form of the 
transformation, 
given by (\ref{2.2a}),  on solving Lie's equations
\ba
\frac{\d \bar{x}}{\d \ve} &=& \xi(\bar{x},\bar{F}, \bar{A})  \n \\
\frac{\d \bar{F}}{\d \ve} &=& \eta(\bar{x},\bar{F}, \bar{A}) \label{bata} \\
\frac{\d \bar{A}}{\d \ve} &=& \zeta(\bar{x},\bar{F}, \bar{A})\n 
\ea
subject to initial conditions
\be \bar{x}\left|_{\ve=0}=x, \qquad \bar{F}|_{\ve=0}=F, \qquad \bar{A}\right|_{\ve=0}=A \label{batas} \ee
The full details on the symmetry approach to solving differential equations can be found in a number of excellent texts 
(Bluman and Kumei 
\cite{blum}, Olver \cite{olver}).

The determination of the generators  is a straight forward process and has been automated by computer algebra 
packages (Dimas and 
Tsoubelis \cite{dimas}, Cheviakov \cite{chev}). In practice, we have found the  package {\tt PROGRAM LIE}  (Head \cite{head}) to be the 
most useful. It is 
quite  accomplished given its age -  it often yields results when its modern counterparts fail.

Utilising {\tt PROGRAM LIE}, we show that (\ref{ben}) admits the following Lie point symmetries/vector fields:
\begin{subequations}
\ba
Z_{1} &=& \frac{\p}{\p x}\label{bobos}\\
Z_{2} &=& x\frac{\p}{\p x}\\
Z_{3} &=& A\frac{\p}{\p A}\\
Z_{4} &=& \frac{F}{n-1}\frac{\p}{\p F}\\
Z_{5} &=& x^2\frac{\p}{\p x}+\frac{xF}{n-1}\frac{\p}{\p F}\label{bobobo}
\ea
\end{subequations}
where $n > 2$. It is  normal practice to use the symmetries (\ref{bobos})-(\ref{bobobo}) to reduce the order of the 
equation in the hope of 
finding solutions of the master equation.  We need to proceed with some caution due to the overdetermined nature of (\ref
{ben}). 
Thereafter we indicate how known solutions can be extended using these symmetries.

\section{New Solutions via Lie symmetries \label{s4}}
One of the main purposes of calculating symmetries is to use them for symmetry reductions and hopefully obtain group 
invariant solutions. 
The goal of this section is to apply the symmetries calculated in Sect.~{\ref{s3} to obtain symmetry reductions and exact solutions 
where possible. 
The application of symmetries (\ref{bobos})-(\ref{bobobo}) to the master equation results in either an implicit solution of 
(\ref{ben}) or we can 
reduce the governing equations to complicated Riccati equations that are difficult to solve. However there are two cases in 
which we can 
find new solutions regardless of the complexity of the function chosen.

\subsection{The choice $A = A(F)$}
An obvious case to consider in this subsection, is when one dependent variable in (\ref{ben}) is a function of the other. 
Usually such an 
approach results in a more complicated equation to solve. In spite of this, we can make significant progress if we use the 
Lie symmetry $Z_
{1}$ (which gives the same result as $Z_{2}$). For our purposes we use the partial set of invariants of
\be
Z_{1} = \frac{\p}{\p x}
\ee
given by
\ba
p &=& F \n \\
q(p) &=& F_{x} \label{mh3} \\
r(p) &=& A \n
\ea
This transformation reduces equation (\ref{ben}) to
\be
q'(p)\left[(n-1)r(p)-pr'(p)\right] = q(p)\left[pr''(p)+2r'(p)\right]\label{she}
\ee
which can be integrated to give
\be
q  =  q_{0}e^{\int{\frac{2r'+pr''}{(n-1)r-r'p}}}dp
\ee
Substituting for the metric functions via (\ref{mh3}),  we can integrate one more time to give the solution
\be
\int\left[e^{-\int\frac{2A_{F}+FA_{FF}}{(n-1)A-FA_{F}}dF}\right]dF  = q_{0}x+x_{0}\label{29}
\ee
where $q_{0}$ and $x_{0}$ are arbitrary functions of time. Equation (\ref{29}) suggests that, given any function $A$ 
depending on arbitrary $F$, we 
can work out $F$ explicitly from (\ref{29}). Such an explicit relationship between $F$ and $A$ has not been found 
previously.  Note that 
since (\ref{ben}) is linear, once we obtain $F$ via (\ref{29}) we can use it to obtain the general solution of (\ref{ben}) using 
standard 
techniques for solving linear equations.

We illustrate this method with simple examples. Using $A = 1$, we evaluate (\ref{29}) to obtain
\be
F = q_{0}(t)x+x_{0}(t)
\ee
We can easily generate the general solution to (\ref{ben}) as
\ba A &=& \frac{-C_1}{q(x_0+qx)}+C_2\\
F &=& \frac{-1+n}{qC_{1}(1+n)}(x_{0}+qx)^{\frac{2}{1-n}}\n \\
&& \mbox{}\left[\frac{1+n}{-1+n}q(x_{0}+qx)^{\frac{1+n}{-1+n}}C_{1}+\left(-C_{1}
+q(x_{0}+qx)C_{2}\right)^{\frac{1+n}{-1+n}}C_{2}\right]
\ea
If we take $A = F^2$, then equation (\ref{29}) is reduced to
\be
\int F^{\left(\frac{6}{3-n}\right)}dF =  q_{0}x+x_{0}
\ee
and hence
\ba
F &=& \left[\frac{9-n}{3-n}(\bar{q}_{0}x+\bar{x}_{0})\right]^{\frac{3-n}{9-n}} \n \\
A &=& \left[\frac{9-n}{3-n}(\bar{q}_{0}x+\bar{x}_{0})\right]^{\frac{6-2n}{9-n}}
\ea
The functional form of $F$ can be easily extended to obtain
\ba
F &=& C_{1}\frac{\left(\frac{9-n}{3-n}\right)^{\frac{3-n}{9-n}}(9-10n
+n^2)(q_{0}x+x_{0})^{\frac{3-n}{9-n}+\frac{-9+2n-n^2}{9-10n+n^2}}}{q_{0}(-9
+2n-n^2)} \n \\&& \mbox{}+C_{2}\left[\frac{9-n}{3-n}(q_{0}x+x_{0})\right]^{\frac{3-n}{9-n}}
\ea
where $C_1$ and $C_2$ are arbitrary functions of 
time, which is the general solution to (\ref{ben}) when
\be A = \left[\frac{9-n}{3-n}(\bar{q}_{0}x+\bar{x}_{0})\right]^{\frac{6-2n}{9-n}} \ee

\subsection{The choice $W = \frac{F}{A^{1/(n-1)}}$}
The combination of symmetries given by
\be
Z_{3}+Z_{4} =  A\frac{\p}{\p A}+\frac{F}{n-1}\frac{\p}{\p F}
\ee
gives rise to the invariant
\be
W  = \frac{F}{A^{1/(n-1)}}\label{mlomos}
\ee
Then equation (\ref{ben}) is transformed by (\ref{mlomos}) to the form
\be
 -nWA_{x}^2+(n-1)A^2W_{xx} = 0 \label{xulu}
 \ee
 with solution
 \be
 A  =   C_1(t)\exp\left({\int\pm \frac{\sqrt{(n-1)^2W_{xx}}}{\sqrt{nW}}dx}\right) \label{swelihle}
 \ee
 which comes as a result of treating equation (\ref{xulu}) as 
 a nonlinear first order ordinary differential equation in $A$.
Given any function $W$ we can integrate the right hand 
side of (\ref{swelihle}) and find a form for $A$.

 If we take $W = a(t)x+b(t)$, then (\ref{swelihle}) gives \be
 A  = \bar{C}_1(t)
 \ee
 and
 \be
  F = {\bar{C}_1(t)}^{1/(n-1)} (a(t)x+b(t))
 \ee
 which are new solutions of (\ref{ben}) for $n>2$.

Alternatively, we could substitute the inverse of (\ref{mlomos}), {\it i.e.}
\be \widehat{W}=\frac{A^{1/(n-1)}}{F}
\ee into (\ref{ben}) and obtain
\be n A_{x}{}^2\widehat{W}^2  - 2(n-1)^2 A^2 \widehat{W}_{x}{}^2 + (n-1)^2A^2\widehat{W}\widehat{W}_{xx} =0 \ee
with solution
\be
A = C_2(t) \exp\left(\pm\int \frac{\sqrt{2(n-1)^{2}\widehat{W}_{x}{}^2 
- (n-1)^2 \widehat{W}\widehat{W}_{xx}}}{\sqrt{n}\widehat{W}}\d x \right) \label{newsol} \ee
Again, given any function $W$ we can integrate 
the right hand side of (\ref{newsol}) and find a form for $A$. 

If we take $W = a(t)x+b(t)$ as before we find that
\be A_1 = C_2(t)\left[a(t)x + b(t) \right]^{\frac{\sqrt{2}(n-1)}{\sqrt{n}}} \ee
and
\be F_1 = \frac{\left[C_2(t)(a(t)x+b(t))^{\frac{\sqrt{2}(n-1)}{\sqrt{n}}}\right]^{\frac{1}{n-1}}}{a(t)x+b(t)} \ee
which is essentially a new solution of the master equation in higher dimensional space. We can also have
\be A_2 = \frac{\bar{C}_2(t)}{(a(t)x+b(t))^{\frac{\sqrt{2}(n-1)}{\sqrt{n}}}}\ee
and
\be F_2 = \frac{\left[\bar{C}_2(t)(a(t)x+b(t))^{\frac{-\sqrt{2}(n-1)}{\sqrt{n}}}\right]^{1/{n-1}}}{a(t)x+b(t)}\ee
thus obtaining two different solutions from the same seed function.

 Observe that  (\ref{swelihle}) and (\ref{newsol}) will contain all 
 solutions of the master equation (\ref{ben}) for appropriately 
 chosen seed functions $W$ or $\widehat{W}$ which are ratios of 
 the metric functions.  We are always able to reduce (\ref{ben})
  to the quadratures (\ref{swelihle}) or (\ref{newsol}) regardless 
  of the complexity of the seed functions.

\section{Extending known solutions \label{s5}}
Another use of Lie point symmetries is the extension of known solutions
 of differential equations.  This is possible due to the fact that the 
 symmetries generate transformations that leave the equations invariant.  
 As a result, applying those transformations to known solutions will (usually) result in new solutions.

We illustrate the approach by using the simple infinitesmal generator $Z_{1}$, where we observe that
\be
\xi = 1, \qquad \eta = 0, \qquad \zeta = 0
\ee
We solve the Lie equations (\ref{bata}), subject to initial conditions (\ref{batas}), to obtain
\ba
\bar{x} &=& x + a_{1}\n \\
\bar{F} &=& F\label{mh1} \\
\bar{A} &=& A \n
\ea
This means that using (\ref{mh1}) we can map the equation (\ref{ben}) to the form
\be
\bar{F}\bar{A}_{\bar{x}\bar{x}}+2\bar{A}_{\bar{x}}\bar{F}_{\bar{x}}-(n-1)\bar{A}\bar{F}_{\bar{x}\bar{x}} = 0 \label{neweq}
\ee
As a result of this mapping, any existing solution to equation (\ref{ben}) 
can be transformed to a solution of (\ref{neweq})  by (\ref{mh1}). 
Usually, $a_{1}$ is an arbitrary constant. However, since $F$ and $A$
 depend on $x$ and $t$ we take $a_{1}$ to be an arbitrary function of time, $a_{1}=a_{1}(t)$.

If we now take each of the remaining symmetries successively, 
we obtain the general transformation
\ba
\bar{x}&=& \frac{e^{a_{2}}(a_{1}+x)}{1-a_{5}e^{a_{2}}(a_{1}+x)} \n \\
\bar{F} &=& \frac{e^{a_{4}}F}{1-a_{5}e^{a_{2}}(a_{1}+x)} \label{mh2}\\
\bar{A} &=& e^{a_{3}}A \n
\ea
where the $a_{i}$ are all arbitrary functions of time and $n > 2$.

Thus any known solution of equation (\ref{ben}) can be transformed 
to a new solution of equation (\ref{ben}) via (\ref{mh2}). 
For example, if we start with the solution
\be
A = 1, \qquad F = \alpha(t)x+\beta(t)
\ee
the transformation (\ref{mh2}) yields the new solution
\ba
\bar{x}&=& \frac{e^{a_{2}}(a_{1}+x)}{1-a_{5}e^{a_{2}}(a_{1}+x)} \n \\
\bar{F}&=& \frac{e^{a_{4}}(\alpha(t) x+\beta(t))}{1-a_{5}e^{a_{2}}(a_{1}+x)}\label{shoz} \\
\bar{A} &=& e^{a_{3}} \n
\ea
All the new results that we derived in the previous section can be similarly extended via (\ref{mh2}).

\section{Conclusion \label{s6}}
In this work, we have provided symmetry reductions and exact 
solutions of the Einstein field equations governing  shear-free 
heat conducting fluids in higher dimensions. Explicit relationships 
were provided between the gravitational potentials, obviating a 
need to start with ``simple'' forms for one to calculate the other.  
We were also able to provide a general transformation to extend 
our (and any other) known solution into new solutions. 
When $n=2$ we regain the results of Deng  \cite{deng} who developed
a method to generate solutions when simple forms of $A$ or $F$ 
are chosen. The case $n=2$ also contains the results of 
Msomi et al.~\cite{msomi} who adopted a more geometric and 
systematic approach using Lie theory to generalise known solutions 
and generate new solutions. The new solutions of this paper may be used to study 
the physics of radiating astrophysical and cosmological models in higher dimensions.

It is interesting to observe an important feature of the solutions 
admitted by equation (\ref{ben}). When $F_{xx}=0$ we find that
(\ref{ben}) becomes 
\[ FA_{xx}+2A_{x}F_{x} =0 \]
This equation has the remarkable feature that it is independent of the dimension $n$.
Thus any solution with $F_{xx}=0$ 
presented by Deng \cite{deng} and Msomi et al.~\cite{msomi} in four dimensions
will also be applicable in higher dimensions. This direct application
of four dimensional solutions into higher dimensional spacetimes
is rather unusual in general relativity.

\begin{acknowledgements}
AMM and KSG thank the National Research Foundation and the 
University of KwaZulu--Natal for ongoing support. AMM also thanks
 Mangosuthu University of Technology for support. SDM acknowledges 
 that this work is based upon research supported by the South African 
 Research Chair Initiative of the Department of Science and Technology 
 and the National Research Foundation.
\end{acknowledgements}

\end{document}